\documentclass[twocolumn,pra, aps,superscriptaddress,showpacs]{revtex4-1}

\usepackage{mathptmx}
\usepackage{subfigure}
\usepackage{dcolumn}
\usepackage{amsmath,amssymb}
\usepackage{bm}
\usepackage{color}
\usepackage{latexsym}
\usepackage{epstopdf}
\usepackage{color}
\usepackage[english]{babel}
\usepackage{latexsym}

\usepackage{psfrag,graphicx} 
\usepackage{epsf} 
\usepackage{subfigure} 
\usepackage{amsmath} 
\usepackage{amssymb} 
\usepackage{amsfonts}
\usepackage{bm}
\usepackage{natbib}
\usepackage{epstopdf}
\usepackage{appendix}

\definecolor{mygrey}{gray}{0.35}
\definecolor{myblue}{rgb}{0.2,0.2,0.8}
\definecolor{myzard}{cmyk}{0,0,0.05,0}
\definecolor{mywhite}{rgb}{1,1,1}
\definecolor{myred}{rgb}{1,0.,0.3}

\usepackage[colorlinks=true,citecolor=myblue,linkcolor=myred]{hyperref}

\def\be{\begin{equation}}
\def\ee{\end{equation}}
\def\ba{\begin{align}}
\def\enda{\end{align}}
\def\bi{\begin{itemize}}
\def\ei{\end{itemize}}

 \def\ee{\mathord{\rm e}}
 
 \def\ii{\mathord{\rm i}}

\def\half{\textstyle\frac{1}{2}}

\def\fourth{\textstyle\frac{1}{4}}

 \def\ee{\mathord{\rm e}}
 
 \def\ii{\mathord{\rm i}}

\def\half{\textstyle\frac{1}{2}}

\def\fourth{\textstyle\frac{1}{4}}

\renewcommand{\ii}{{\rm i}}
\renewcommand{\ee}{{\rm e}}

\def\beq{\begin{equation}}
\def\beq{\begin{equation}}
\def\eeq{\end{equation}}

 \newcommand{\ket}[1]{|#1\rangle}
 \newcommand{\bra}[1]{\langle #1|}

\begin{document}

\title[Short Title]{Electron-Mediated Nuclear-Spin Interactions Between Distant NV Centers}

\author{A. Bermudez}
\affiliation{Institut f\"{u}r Theoretische Physik, Albert-Einstein Allee 11, Universit\"{a}t Ulm, 89069 Ulm, Germany}
\author{F. Jelezko}
\affiliation{Institut f\"{u}r Quantenoptik, Albert-Einstein Allee 11, Universit\"{a}t Ulm, 89069 Ulm, Germany}
\author{M. B. Plenio}
\affiliation{Institut f\"{u}r Theoretische Physik, Albert-Einstein Allee 11, Universit\"{a}t Ulm, 89069 Ulm, Germany}
\author{A. Retzker}
\affiliation{Institut f\"{u}r Theoretische Physik, Albert-Einstein Allee 11, Universit\"{a}t Ulm, 89069 Ulm, Germany}


\begin{abstract}
{We propose a scheme enabling controlled quantum coherent interactions between separated nitrogen-vacancy
centers in diamond in the presence of strong magnetic  fluctuations. The proposed scheme couples nuclear
qubits employing the magnetic dipole-dipole interaction between the electron spins and, crucially, benefits from the
suppression of the effect of environmental magnetic field fluctuations thanks to a strong microwave driving.
This scheme provides a basic building block for a full-scale quantum information processor or quantum simulator
based on solid-state technology. }
\end{abstract}

\maketitle


The spins of single dopants in solids are key elements in the development of solid-state quantum-information  technologies~\cite{qc_defects,qc_review}. In particular,  nitrogen-vacancy (NV) colour centers in diamond are promising quantum processors:  single defects can be detected using confocal microscopy~\cite{nv_confocal,single_nv}, their spin state can be initialized, manipulated, and readout optically~\cite{nv_optical_pumping,nv_electron_rabi,single_spin_spectrosocpy,c13_environment}, and their quantum coherence survives at room temperatures~\cite{long_coherence_time}. One of the remaining  challenges is to control the spin-spin interactions to perform quantum-logic operations, and major steps along this direction have already been accomplished. The hyperfine coupling between the NV electron spin and the nuclear spins of neighboring impurities ($^{13}$C,$^{15}$N)  offers a unique opportunity to build small quantum registers~\cite{nv_n_coupling,nv_nuclear_single_shot,c13_entanglement,c13_environment,c13_register}. These devices can be  scaled up by means of ion implantation techniques, yielding  periodic arrays of NV centers~\cite{cold_ion_implantation}. However, the controlled  couplings now require longer-range interactions, as provided by optical channels~\cite{nv_photon}, or magnetic dipole-dipole couplings between the electron spins~\cite{nv_nv_gate}. 

Although  the feasibility of the magnetic-coupling approach has been  demonstrated recently~\cite{nv_nv_gate}, fabricated NV arrays often suffer from shorter electron coherence times that affect  the fidelity of the quantum gates. From this perspective,   $^{14}$N or   $^{15}$N  nuclear spins would be better-suited qubits due to their longer coherence times, together with the availability of single-shot readout~\cite{nv_nuclear_single_shot}. Unfortunately, the direct nuclear dipole-dipole interaction is negligible, which necessitates the search for alternative schemes to  couple the nuclear spins. This letter presents a theoretical proposal for implementing robust quantum gates between two distant nuclear-spin qubits mediated by the long-range dipolar interaction between electron spins. The main idea is to exploit the long nuclear coherence times for storage, and to use the electronic degrees of freedom as a quantum bus that mediates the nuclear spin interaction. Such a general  scheme can be applied to different setups, and has also been proposed for quantum-Hall systems~\cite{Hall}. Active control of the spins via microwave fields allows reaching high fidelities, even in the presence of the magnetic noise associated to the complex mesoscopic environment of solid-state systems. In fact, the nuclear driving acts as a continuous decoupling mechanism~\cite{cont_decoupling} that minimizes the effects of the  noise, and provides a new tool in addition to pulsed  techniques~\cite{dynamical_decoupling_nv}.


{\it The model.-} We consider two NV defects $j=1,2$,  whose unpaired electrons  form a spin-triplet  ground state $S_j=1$, and focus on $^{14}$N with  a nuclear spin $I_j=1$. The  Hamiltonian that describes each  NV center is $H_j=H_j^{\text{(e)}}+H_j^{\text{(n)}}+H_{j}^{\text{(e-n)}},$
 \begin{equation}
 \label{local_hamiltonian}
 \begin{split}
 H_j^{\text{(e)\phantom{()}}}&=D_j\left((S_j^z)^2-\textstyle{\frac{1}{3}}\boldsymbol{S}_j^2\right)+ g_{\text{e}}\mu_{\text{B}}\boldsymbol{B}\cdot\boldsymbol{S}_j,\\
  H_j^{\text{(n)\phantom{()}}}&=-P_j\left((I_j^z)^2-\textstyle{\frac{1}{3}}\boldsymbol{I}_j^2\right)- g_{\text{n}}\mu_{\text{N}}\boldsymbol{B}\cdot\boldsymbol{I}_j,\\
   H_j^{\text{(e-n)}}&= A^{\shortparallel}_jS_j^zI_j^z+\half A^{\bot}_j(S_j^{+}I_j^-+S_j^{-}I_j^+),\\
 \end{split}
 \end{equation}
where $\boldsymbol{S}_j,\boldsymbol{I}_j$ are the electronic and nuclear spin-1 operators, and  ${S}_j^{\pm}=S^x_j\pm \ii S^y_j$, ${I}_j^{\pm}=I^x_j\pm \ii I^y_j$ the usual ladder operators. Here, $D_j$($P_j$) stands for the zero-field splitting of the electronic (nuclear) ground state, $\boldsymbol{B}$ is an external magnetic field, $\mu_{\rm B} (\mu_{\rm N})$ is the Bohr (nuclear) magneton, and $g_{\rm e}(g_{\rm n})$ is the electron (nuclear) g-factor. The electron-nuclei interaction  is quantified by the hyperfine longitudinal (transverse) coupling $A_j^{\shortparallel}$ ($A_j^{\bot}$). The present discussion is focused on a single pair of closely-spaced NV centers, and we use  the realistic parameters of the experiment in~\cite{nv_nv_gate}. We emphasize, however, that this scheme can be extended to arrays of implanted NV centers, provided that their  distance is small enough. Let us also remark the hierarchy of couplings,  $D_j\gg P_j\gtrsim A_j^{\shortparallel},A_j^{\bot}$, and $g_{\text{e}}\mu_{\text{B}}\gg g_{\text{n}}\mu_{\text{N}}$ 
(see Table~\ref{o_magnitude}, where  $\hbar=1$). Finally, we  introduce the  secular dipole-dipole interaction between the electron spins 
\begin{equation}
\label{dipole_hamiltonian}
H_{12}^{\text{(e-e)}}=J_{12}\left(3S_1^zS_2^z-\boldsymbol{S}_1\cdot \boldsymbol{S}_2\right),
\end{equation}
where $J_{12}=g_{\text{e}}^2\mu_{\text{B}}^2(1-3\cos^2\theta_{12})/2 c r_{12}^3$ in gaussian units, $\boldsymbol{r}_{12}$ is the distance between the NV centers, $\cos \theta_{12}=\boldsymbol{e}_{z}\cdot \boldsymbol{r}_{12}/r_{12}$, and $c$ is the speed of light.  For the distances reached in the experiment, $r_{12}\approx10$nm, the dipolar coupling $J_{12}\approx 70\text{kHz}$ is the smaller energy scale in the problem. As mentioned above, the magnetic dipole-dipole interaction between the nuclear spins is completely negligible since $(g_{\text{n}}\mu_{\rm N}/g_{\text{e}}\mu_{\rm B})^2\approx 10^{-8}$, and an indirect mechanism for the  nuclear coupling  is thus required.

\begin{figure}
	\centering
	\includegraphics[width=1\columnwidth]{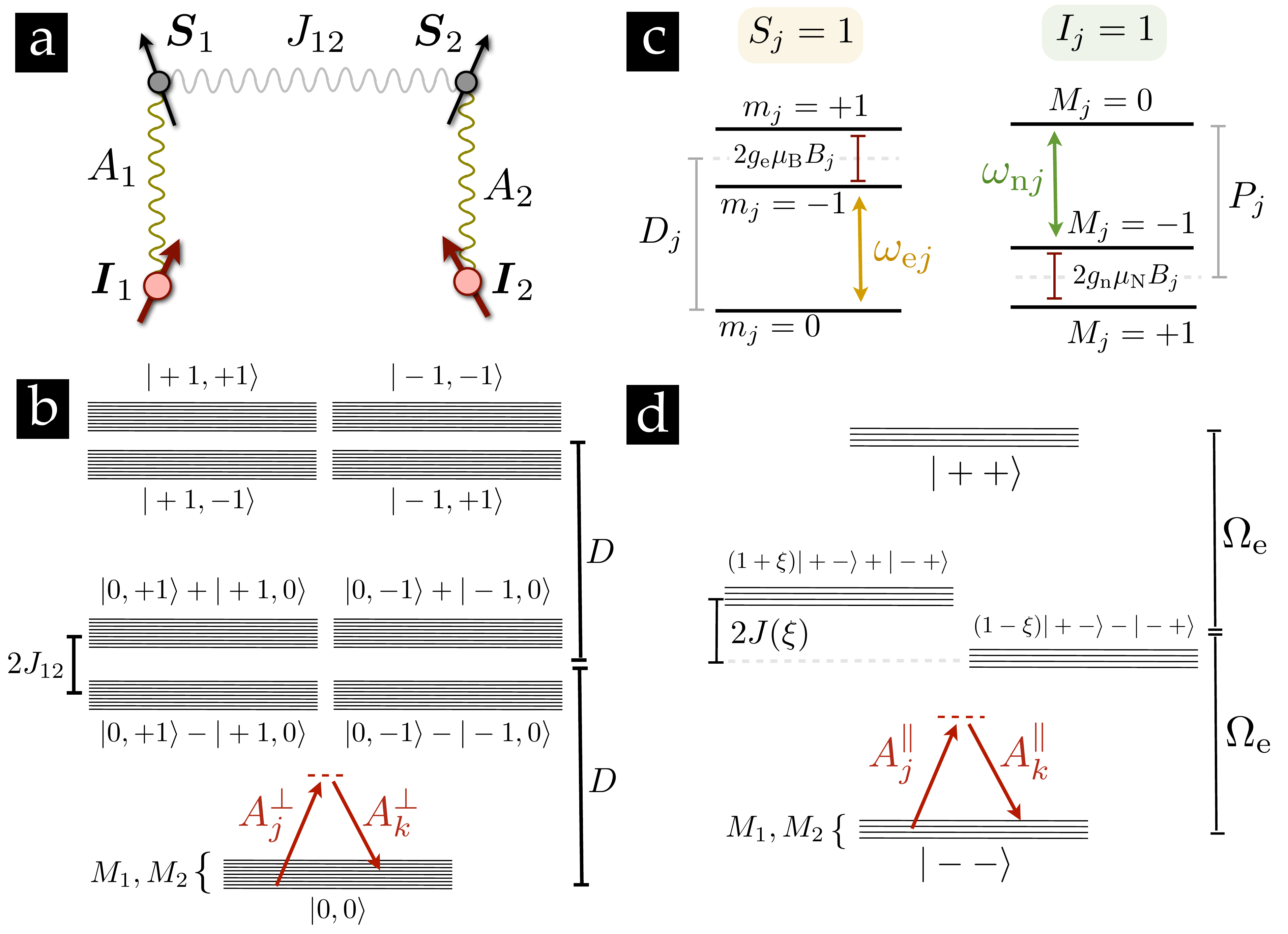}
	\caption{\label{cones} {\bf Effective nuclear spin-spin interaction.} (a) Schematic diagram of the electron-mediated interaction between the nuclear spins, which  exploits the magnetic dipolar interaction and the local hyperfine coupling. (b) Diagram of the energy levels of the two NV centers. Since $D_j$ is the largest energy scale, the energies are clustered in manifolds determined by the electronic spin. The transverse part of the hyperfine coupling $A_{j}^{\bot}$ induces transitions between different manifolds, and  mediates an effective XX interaction between the nuclear spins. (c) Schematic diagram of the Zeeman splitting for the electronic and nuclear energy levels. By carefully selecting the microwave frequencies,  we drive a particular electronic and nuclear transition.  (d) Energy levels of the driven Hamiltonian. For very strong driving $\Omega_{\text{e}}$, the  electronic spin in the lowest manifold is $\ket{-}\propto\ket{0}-\ket{{\rm -1}}$. The hyperfine coupling $A_{j}^{\shortparallel}$ induces virtual transitions to the excited manifold, split by the dipolar interaction and the inhomogeneous broadening, and leads to an effective ZZ  interaction.  }
	\label{effective_coupling}
\end{figure}

{\it Effective static interactions.-} In Fig.~\ref{effective_coupling}{\bf(a)}, we represent  schematically  the process leading to nuclear spin-spin interactions. The  hyperfine interaction couples the nuclear to the electronic spins of each NV center, which are in turn coupled through the magnetic dipole-dipole interaction. Therefore, one may use the electrons as a bus to mediate the nuclear coupling. A naive estimate of this coupling  follows from Fig.~\ref{effective_coupling}{\bf(b)}, where we represent the energy spectrum of $H_0=\sum_{j}(H_j^{\text{(e)}}+H_j^{\text{(n)}})+H_{12}^{\text{(e-e)}}$. Due to the energy-scale hierarchy in Table~\ref{o_magnitude}, the  levels are clustered in manifolds determined by the electronic spins $\ket{m_1,m_2}_{\rm e}$. The dynamics within the ground-state manifold, $\ket{0,0}_{\rm e}$, corresponds to nuclear spin flips $\ket{M_1,M_2}_{\rm n}\to\ket{M'_1,M'_2}_{\rm n},$ with $M_j,M'_j=0,\pm 1$, and follows from second-order processes where the hyperfine coupling virtually populates states from the excited manifold. Therefore, a crude estimate of the dynamics is $H_{\text{eff}}\approx J_{\text{eff}}I_1^+I_2^-+\text{H.c.}$, where $J_{\text{eff}}\propto(A_1^{\bot}A_2^{\bot})/D$. A more careful Schrieffer-Wolff-type calculation  takes into account the two possible channels, symmetric or anti-symmetric, which lead to the destructive interference of this coupling  $J_{\text{eff}}\propto (A_1^{\bot}A_2^{\bot})/D-(A_1^{\bot}A_2^{\bot})/D$. It is precisely the role of the magnetic dipole-dipole interaction to split these channels, suppressing the perfect destructive interference, and leading to 
\begin{equation}
\label{eff_static}
H_{\rm eff}^{\rm xx}=J_{\rm eff}^{\rm xx}(I_1^+I_2^-+I_1^-I_2^+)-\hspace{-0.5ex}\sum_{j}P_j(I_j^z)^2,\hspace{0.5ex}J_{\text{eff}}^{\rm xx}=\frac{2A_1^{\bot}A_2^{\bot}}{D^2}J_{12}.
\end{equation}
This Hamiltonian describes the {\it flip-flop  interaction} between the $^{14}$N nuclei leading to an exchange of the spin excitations. 

\begin{table*}
  \centering 
   \caption{{\bf Specific values of the coupling strengths}  }
  \begin{tabular}{ c  c c  c  c  c c  c c  c  c c c}
\hline
\hline
$D_j$ & $P_j$ & $A_j^{\shortparallel},A_j^{\bot}$ & $J_{12}$ & $g_{\text{e}}\mu_{\text{B}}$ & $g_{\text{n}}\mu_{\text{N}} $& $B $ &  $\Omega_{\text{e}}$ & $\Omega_{\text{n}}$ & $J^{\rm xx}_{\text{eff}}$ & $J^{\rm zz}_{\text{eff}}$ \\
\hline
\hspace{0.5 ex}2.87 GHz\hspace{0.5ex} &\hspace{0.5ex} 5.04 MHz \hspace{0.5ex}& \hspace{0.5ex}2.1,2.3 MHz\hspace{0.5ex}  &\hspace{0.5ex} 70 kHz\hspace{0.5ex} & \hspace{0.5ex}2.8 MHz$\cdot$ G$^{-1}$\hspace{0.5ex} & \hspace{0.5ex} 0.31 kHz$\cdot$ G$^{-1}$ \hspace{0.5ex} & \hspace{0.5ex} 30 G \hspace{0.5ex} & \hspace{0.5ex} 15 MHz \hspace{0.5ex} & \hspace{0.5ex} 1 kHz \hspace{0.5ex}  & \hspace{0.5ex} 0.1 Hz \hspace{0.5ex} & \hspace{0.5ex} 0.1 kHz \hspace{0.5ex} \\
\hline
\hline
\end{tabular}
  \label{o_magnitude}
\end{table*}

\begin{figure*}
	\centering
	\includegraphics[width=2\columnwidth]{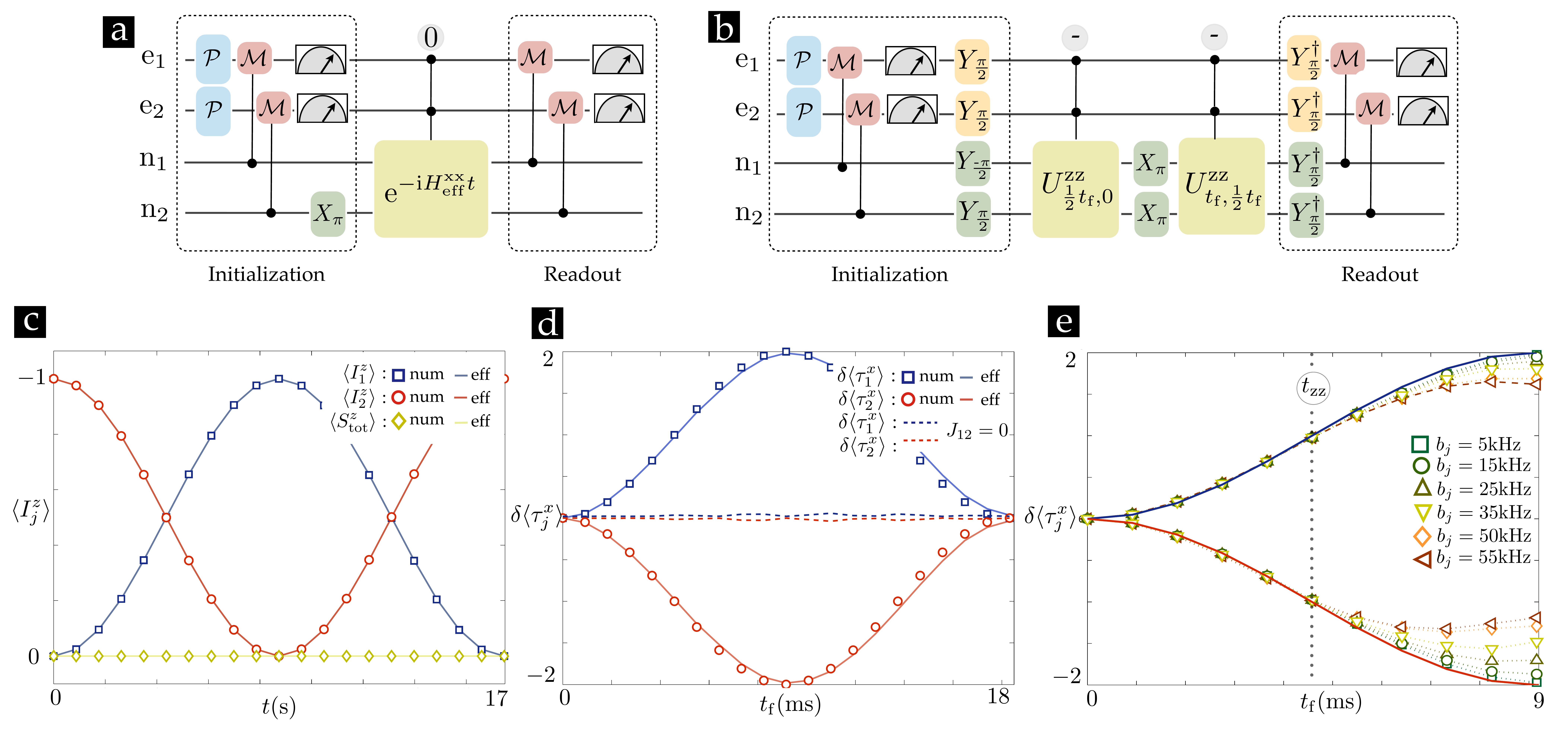}
	\caption{{\bf Nuclear quantum gate between two NV centers:}  Scheme for the initialization, evolution, and read-out of the effective $J^{\rm xx}_{\rm eff}$ interaction in (a), and $J^{\rm zz}_{\rm eff}$ interaction in (b). Here,   $X_{j,\phi}={\rm exp}({\ii\phi\tau^x_j}/2),Y_{j,\theta}={\rm exp}(\ii\theta\tau_j^y/2)$ represent finite pulses, and $\mathcal{P}$($\mathcal{M}$)  stand for the electron (nuclear) spin polarization. (c) Comparison between the effective dynamics under the Hamiltonian~\eqref{eff_static} and the exact time evolution under Eqs.~\eqref{local_hamiltonian}-\eqref{dipole_hamiltonian} for the $J^{\rm xx}_{\rm eff}$  nuclear interaction.  The expectation values represented correspond to the nuclear spin $\langle I_j^z\rangle$, and electronic spin $\langle S_{\rm tot}^z\rangle=\langle S_{1}^z+S_{2}^z\rangle$. (d) Comparison between the exact~\eqref{driven_hamiltonian} and effective~\eqref{ising} dynamics for the $J^{\rm zz}_{\rm eff}$ nuclear interaction, together with an echo scheme that allow us to get rid of the fast single-nuclei dynamics. We represent the nuclear expectation values $\delta\langle \tau_j^x\rangle=\langle \tau_j^x\rangle_{t_{\rm f}}-\langle \tau_j^x\rangle_{t_{\rm 0}}$. The dotted lines correspond to $J_{12}=0$, where there is no interaction induced on the nuclei. (e) Performance of the ZZ-gate in the presence of different strengths of the electron dephasing noise $b_j=\{5,15,25,35,50,55\}$kHz, where the nuclear noise is $B_j=0.1b_j$. The corresponding Ramsey decoherence times are roughly $T_{2{\rm e}}\approx\{0.2,0.07,0.04, 0.03, 0.02,0.018\}$ms, $T_{2{\rm n}}=0.1T_{2{\rm e}}$. }
	\label{nuclear_gate}
\end{figure*}

In Fig.~\ref{nuclear_gate}{\bf(a)}, we present a scheme for the electron-mediated  gate between two NV nuclei based on Eq.~\eqref{eff_static}, referred as the nuclear XX gate. The initialization  yields the  state $\ket{\psi_0}=\ket{\phi_{\rm e}}\otimes\ket{\varphi_{\rm n}}=\ket{0,0}_{\rm e}\otimes\ket{0,1}_{\rm n}$, where electrons belong to the  ground-state manifold of Fig.~\ref{effective_coupling}{\bf(b)}, and  the dynamics of the spin excitation is determined by virtual electron spin-flip processes. In Fig.~\ref{nuclear_gate}{\bf(c)}, we study numerically the  accuracy of the effective Hamiltonian~\eqref{eff_static}, which is compared to the exact evolution under the total Hamiltonian~\eqref{local_hamiltonian}-\eqref{dipole_hamiltonian}. One  observes that the electron state remains in the ground-state, whereas there is a periodic exchange of the spin excitation between the nuclei. The remarkable agreement of both predictions justifies the validity of the effective nuclear spin-spin Hamiltonian in Eq.~\eqref{eff_static}. Unfortunately, the parameters in Table~\ref{o_magnitude} yield a vanishingly-small   coupling  $J^{\rm xx}_{\text{eff}}\approx 0.1$Hz, which is far  too slow to produce any observable coherent coupling between the nuclei. Even if not of  practical use, the above derivation gives a neat account of the mechanism of electron-mediated interactions, and will help us in understanding how to raise the interaction strength.

A possibility to overcome this problem is to apply a magnetic field, such that  the Zeeman shift  reduces  $D\to D-g_{\text{e}}\mu_{\text{B}}B$, thus enhancing $J^{\rm xx}_{\text{eff}}$. Yet, one faces two important problems: {\it i)} In general, the axes of the NV centers are not aligned, and  each electronic spin experiences a different Zeeman shift. For the large fields required, this inhomogeneity might exceed the dipolar coupling, and thus spoil the scheme. {\it ii)} The dephasing  exerted by the environment  would have a contribution  that ruins the coherence of the  interaction.  We demonstrate below  that there is a different approach that  overcomes both problems simultaneously, and yet enhances the  nuclear spin interaction: {\it continuous microwave driving}~\cite{cont_decoupling}.

{\it Effective driven interactions.-}  We discuss now the effects of a continuous microwave field that drives both the electronic and nuclear spins. The effect of the driving is two-fold: {\it i)} By addressing each NV center with different microwave fields, one can independently tune their frequencies so that they become resonant with a particular transition. This allows us to overcome the problems associated with both the inhomogeneous broadening, and the different Zeeman shifts. Moreover, this can be used for single addressing of NV's, especially when combined with magnetic gradients. {\it ii)} By tuning the microwave frequency  on resonance with the transition, one introduces a new energy scale that governs the system, namely the Rabi frequency. This parameter can be tuned by controlling the microwave power, allowing us to enhance $J_{\text{eff}}$.

 Let us  consider the   Zeeman effect  associated to $B=30$ G in Fig.~\ref{effective_coupling}{\bf(c)}. By setting the microwave frequencies to  $\omega_{\text{e}j}=D_j-g_{\text{e}}\mu_{\text{B}}B_j, \omega_{\text{n}j}=P_j-g_{\text{n}}\mu_{\text{N}}B_j$, one resonantly drives  the transitions between the electronic and nuclear  levels  $m_j=0\leftrightarrow-1$, $M_j=0\leftrightarrow -1$. These driving terms  can be written as
\begin{equation}
\label{driving}
H_{{\rm d}}(t)=\sum_{j}\Omega_{\text{e}}\sigma^x_j\cos{ \omega_{\text{e}_j} t}+\Omega_{\text{n}}\tau^x_j\cos{ \omega_{\text{n}_j} t},
\end{equation}
where  the Rabi frequencies  of the electronic and nuclear transitions are $\Omega_{\text{e}},\Omega_{\text{n}}$, and the  electronic and nuclear Pauli matrices  $\sigma_j^x$, $\tau_j^x$.
In the interaction picture with respect to $H_{0,1}=\sum_{j}D_j(S_j^z)^2-P_j(I_j^z)^2+g_{\text{e}}\mu_{\text{B}}B_jS_j^z-g_{\text{n}}\mu_{\text{N}}B_jI_j^z$, one can neglect the rapidly oscillating terms associated to the transverse part of Zeeman shifts, and the hyperfine coupling. This rotating wave approximation is justified  for the parameters shown in Table~\ref{o_magnitude}. Additionally, we  consider two NV centers with  different axes, which  allows us to neglect the transverse part of the magnetic dipole coupling. For weak-enough driving, we arrive at the the total {\it driven Hamiltonian}
\begin{equation}
\label{driven_hamiltonian}
H_0=\hspace{-0.5ex}\sum_{j}\hspace{-0.5ex}\left(\half \Omega_{\text{e}}\sigma^x_j+\half\Omega_{\text{n}}\tau^x_j\right)+2J_{12}S_1^zS_2^z,\hspace{1ex}H_1\hspace{-0.5ex}=\hspace{-0.5ex}\sum_{j}A_j^{\shortparallel}S_j^zI_j^z.\\
\end{equation}
We stress that  these approximations are justified by the  parameters in Table~\ref{o_magnitude}, and supported by  numerical simulations.

We  derive now the electron-mediated nuclear spin interactions starting from  Eq.~\eqref{driven_hamiltonian}. We note that there is again a hierarchy in the  couplings $\Omega_{\text{e}}\gg A_j^{\shortparallel}\gg J_{12}\gg \Omega_{\text{n}}$, which leads to the clustering of energy levels shown in Fig.~\ref{effective_coupling}{\bf(d)}. By considering the electron ground-state, the nuclear spins can interact through virtual electron spin-flips to the excited manifolds. In this driven regime, it is the  longitudinal hyperfine coupling $A_j^{\shortparallel}$ which induces such virtual transitions. A Schrieffer-Wolff-type calculation yields the nuclear Hamiltonian 
\begin{equation}
\label{ising}
H_{\text{eff}}^{\rm zz}\hspace{-0.5ex}=J^{\rm zz}_{\text{eff}}\tau_1^z\tau_2^z+\hspace{-0.5ex}\sum_j\hspace{-0.5ex}\Omega_{\text{n}}\tau_j^x-\fourth A_j^{\shortparallel}\tau_j^z,\hspace{0.5ex} J^{\rm zz}_{\text{eff}}=\hspace{-0.5ex}\frac{-A_1^{\shortparallel}A_2^{\shortparallel}}{8\Omega_e}\hspace{-0.5ex}\left(\hspace{-0.5ex}\frac{J_{12}}{\Omega_{\text{e}}}+2\xi\hspace{-0.5ex}\right)\hspace{-0.5ex},
\end{equation}
where we considered  the inhomogeneous broadening of the hyperfine couplings $ \xi=2[(A_2^{\shortparallel})^2-(A_1^{\shortparallel})^2]/\Omega_eJ_{12}$. This Hamiltonian is an {\it Ising magnetic interaction} between the nuclear spins, which are additionally subjected to a { transverse field} due to the driving, and a { longitudinal field} due to the hyperfine coupling. As advanced previously, we have been able to enhance the electron-mediated nuclear interaction, which becomes $J^{\rm zz}_{\text{eff}}\approx 0.1$kHz for the parameters in Table~\ref{o_magnitude}. Remarkably, the strength of the nuclear spin interaction has increased by three orders of magnitude $J^{\rm zz}_{\text{eff}}\approx 10^3J^{\rm xx}_{\text{eff}}$.

In Fig.~\ref{nuclear_gate}{\bf(b)}, we schematically describe the necessary ingredients for the nuclear ZZ  gate. The initialization consists of the electron (nuclear) spin polarization $\mathcal{P}$($\mathcal{M}$), together with single-spin gates.  $\mathcal{P}$ is obtained by the  optical pumping cycle available for NV centers~\cite{nv_optical_pumping,nv_electron_rabi}, whereas $\mathcal{M}$ is  based on the techniques developed for the nuclear single-shot measurement~\cite{nv_nuclear_single_shot}, followed by the  electron state-dependent fluorescence~\cite{nv_optical_pumping,nv_electron_rabi}.  Once polarized, $\ket{0,0}_{\rm e}\otimes\ket{0,0}_{\rm n}$, one  applies unitary gates based on microwave pulses of different duration, $
Y_{j,\frac{\pi}{2}}=(\mathbb{I}+\ii\tau_j^y),\hspace{1ex} Y_{j,-\frac{\pi}{2}}=(\mathbb{I}-\ii\tau_j^y)$ (also for the electron spin), which lead to $\ket{\psi_0}=\ket{{\rm --}}_{\rm e}\otimes\ket{{\rm -+}}_{\rm n}$.  The evolution of this state is dictated by the interaction-picture Hamiltonian~\eqref{ising}, which leads to $U^{\rm zz}_{t_2,t_1}=\ee^{-\ii H_{0,1}t_2}\ee^{-\ii H_{\rm eff}^{\rm zz}(t_2-t_1)}\ee^{+\ii H_{0,1}t_1}$. Due to the longitudinal field, and the additional contributions of  $H_{0,1}$, the simple periodic exchange of the nuclear spin excitation shall be accompanied by fast oscillations.  In order to observe neatly the effect of the interaction, one may perform a spin-echo sequence, such that the nuclear spins are inverted at half the gate time by a microwave pulse $X_{j,\pi}=\ii\tau_{j}^x$. In this case, the fast single-nuclei oscillations refocus after the spin-echo period $t_{\rm f}$, and one observes solely the effect of the interaction. In Fig.~\ref{nuclear_gate}{\bf(d)},
we compare the effective description~\eqref{ising} to the Hamiltonian~\eqref{driven_hamiltonian}, which display a clear agreement. In particular, when the echo period matches twice the ZZ-gate time $t_{\rm f}=2t_{\rm zz}=\pi/2J_{{\rm eff}}^{{\rm zz}}\approx 9$ms, one finds  a perfect excitation exchange  $\langle \tau^x_1\rangle:-1\to+1,\langle\tau^x_2\rangle:+1\to-1$. Note that for $J_{12}=0$, this effect is completely absent. Finally, considering $t_{\rm f}=t_{\rm zz}$, and setting the echo pulse along the y-axis, the dynamics generates a  entangled nuclear state $\ket{\psi_0}=\ket{{\rm --}}_{\rm e}\otimes\ket{{\rm -_y+_y}}_{\rm n}\to\ket{\psi_{\rm f}}=\ket{{\rm --}}_{\rm e}\otimes(\ket{{\rm -_y+_y}}_{\rm n}+\ket{{\rm +_y-_y}}_{\rm n})/\sqrt{2}$. Once the gate has been performed,  the nuclear  operators $\langle I_j^z\rangle$, $\langle \tau_j^x\rangle$ must be measured. Since the state-dependent fluorescence is particular to the electron spins, one should map the nuclear information onto the electrons, and then measure. This can be achieved  in a quantum non-demolition fashion by using a microwave  on a electron-spin transition conditioned to the nuclei~\cite{nv_nuclear_single_shot}.

{\it Decoupling from decoherence.-} So far, our discussion  has focused on the idealized  situation of isolated NV centers. However, every quantum system is inevitably coupled to an environment that degrades its coherence. This phenomenon, known as {\it decoherence}, must be seriously accounted for in solid-state materials,
where the  system-environment coupling is usually strong. In the particular case of NV centers, the major source of decoherence is the coupling  to
other impurity spins, such as single substitutional  nitrogen electron spin (P1 center)  in type Ib diamond~\cite{adjustable_spin_bath_nv}, or  $^{13}$C isotopes in type IIa~\cite{c13_environment}. The microscopic description of the spin bath is an intricate many-body problem, and is a current subject of intense  research. Here, we use a phenomenological  model of the bath that yields a fluctuating  magnetic field  shifting the resonance frequencies. Due to the spin interactions, this effective field is modeled as a stochastic Ornstein-Uhlenbeck process~\cite{dynamical_decoupling_nv,rabi_oscillations_decay}
\begin{equation}
\label{noise}
H_{\rm noise}=\sum_j \left( b_{j}(t)S_j^z+ B_{j}(t)I_j^z\right),
\end{equation}
where $b_{j}(t), B_{j}(t)$ are random processes with autocorrelation $
\langle b_j(t)b_j(0)\rangle = b_j^2\ee^{-r_j t}, \hspace{1ex} \langle B_j(t)B_j(0)\rangle =B_j^2\ee^{-R_j t},$
where $b_j^2,B_j^2$ represent variance of the zero-mean gaussian distributions, and $r_j,R_j$ the inverse of their correlation times. In particular, the decoherence time of an electronic (nuclear) Ramsey experiment is given by $T_{2\rm e}=1/b_j$ ($T_{2\rm n}=1/B_j$). By considering the particular time-dependence of these stochastic processes, we numerically integrate the noisy dynamics, and average for  $N=10^3$ realizations of the random process. This allows us to study the effects of decoherence on the gate.

For the slow XX gate (Fig.~\ref{nuclear_gate}{\bf(a)}), the limiting factor is the nuclear dephasing time, which can attain values of $T_{2{\rm n}}\approx 10{\rm ms}$. Even for the purest samples, the coherence of the gate is completely lost much before the target time $t_{\rm xx}\approx 4.5$s is reached. Therefore, the  performance of this gate is extremely poor. For the fast ZZ gate (Fig.~\ref{nuclear_gate}{\bf(b)}), not only the nuclear-spin dephasing, but also the electron-spin dephasing limit the gate accuracy. In the dressed-state basis (see Fig.~\ref{effective_coupling}{\bf(d)}), the electron dephasing tries to induce a transition between the different manifolds, introducing additional noise in the nuclei. However, due to the strong driving $\Omega_{\rm e}$, these processes are partially suppressed. Additionally, a sufficiently strong nuclear driving, $\Omega_{\rm n}\gg B_j,(b_jA_j^{\shortparallel}/\Omega_{\rm e})$,  provides an additional decoupling mechanism that enhances further the gate performance. In Fig.~\ref{nuclear_gate}{\bf(e)}, one observes the announced decoupling, since  the gate performance at the target time $t_{\rm zz}\approx 4.5$ms is extremely good even for  shorter electronic coherence times ranging from $T_{2{\rm e}}\approx 0.1$ms to $T_{2{\rm e}}\approx 50\mu$s.  Due to the decoupling mechanism, the gate accuracy will actually be limited by the decay times $T_{1\rm{e}}$. Moreover, at this time scale, energy will be pumped into the system by the continuous driving. However, note that this limitation can be overcome since $T_{1{\rm e}}$ can be increased by orders of magnitude by  cooling. Accordingly, one can achieve high fidelities.

Let us finally note that the effective decoupling mechanism presented here can also be used to improve the electron-spin gates based on the direct dipole interaction~\cite{nv_nv_gate}. In that case, the role of the microwave driving is to prolong dephasing times and to bring the two dressed electronic transitions to resonance to overcome the inhomogeneous broadening.

{\it Conclusions and outlook.-} We have demonstrated the feasibility for engineering electron-mediated spin-spin interactions between
the nuclei of two NV-centers. By continuous microwave driving, this scheme allow us to decouple from
the electronic and nuclear dephasing sources, and increase the effective interactions by three orders of
magnitude magnitude thus achieving $J_{\rm eff}^{\rm zz}\approx 0.1$kHz for distances of existing pairs of NV-centers~\cite{nv_nv_gate}. This scheme opens the possibility
for the realization of quantum information processors, quantum simulators and quantum-sensors~\cite{nv_magnetrometry} on the
basis of NV-centers in diamond. Finally, we would like to stress the generality of this scheme, which can be applied to other solid-state technologies that are candidates for quantum-information processing.

{\it Acknowledgements.--} This work was supported by the EU STREP projects
HIP, PICC, and by the Alexander von Humboldt Foundation. We thank J. Cai for useful discussions.

\vspace{-4ex}


\newpage

\begin{widetext}
\section{Supplementary material}

In the following sections, we provide a detailed discussion of some technical aspects used to derive the main results presented above. In the first section, we derive the effective nuclear Hamiltonians in Eqs.~\eqref{eff_static}-\eqref{ising} by means of the so-called Schrieffer-Wolff transformation. Finally, in the second section, we describe the phenomenological model used to study the effects of  the decoherence on the nuclear-gate performance.  

\appendix

\vspace{1ex}

\appendix

\subsection{Schrieffer-Wolff transformation and effective nuclear Hamiltonians}
\label{sw}

In this section, we review the theory of quasi-degenerate perturbation theory as a tool to perform the so-called {\it adiabatic elimination} of the fast degrees of freedom in a quantum-mechanical system~\cite{cohen_atom_photon}. In particular, this technique allows us to derive the effective spin-spin Hamiltonians for the nuclei.

{\it Schrieffer-Wolff transformation.-}We shall assume that the Hamiltonian is of the form $H=H_0+\lambda V$, where $\lambda V$ is a weak perturbation to the unperturbed Hamiltonian $H_0=H_{{A}}+H_{{B}}$. Here, $H_{A}$ and $H_{B}$ describe the {\it slow} and {\it fast} degrees of freedom  
\begin{equation}
H_{A}\ket{{j\alpha}}=E_{\alpha}\ket{{\alpha\beta}},\hspace{2ex}H_{B}\ket{{\alpha\beta}}=E_{\beta}\ket{{\alpha\beta}},
\end{equation} where $|E_{\alpha}-E_{\alpha'}|\ll |E_{\beta}-E_{\beta'}|$. Accordingly, the frequencies associated to the $\beta\to\beta'$ transitions, $\omega_{\beta\beta'}=E_{\beta}-E_{\beta'}$, are much larger than those of $\alpha\to \alpha'$, $\omega_{\alpha\alpha'}=E_{\alpha}-E_{\alpha'}$, and they can be adiabatically eliminated.  By performing a canonical transformation $U=\ee^{S}$,  $S^{\dagger}=-S$, one constructs an effective Hamiltonian $H_{\text{eff}}=U^{\dagger}HU$  that only involves the slow degrees of freedom, $P_{\beta}H_{\text{eff}}P_{\beta'}=H_{\text{eff}}^{\beta}\delta_{\beta\beta'}$, where $P_{\beta}=\sum_{\alpha}\ket{{\alpha\beta}}\bra{{\alpha\beta}}$. The canonical transformation $S$, and the effective Hamiltonian $H_{\text{eff}}$, can be constructed to any order of the perturbative parameter $\lambda$. This is usually known as the Schrieffer-Wolff transformation in condensed matter~\cite{schrieffer_wolff}, although it also arises in a broader context~\cite{winkler,previous_biblio}. To second order~\cite{cohen_atom_photon}, one finds the following expression
\begin{equation}
\label{effective_ham}
\bra{\alpha}H_{\text{eff}}^{\beta}\ket{\alpha'}\hspace{-0.25ex}=\hspace{-0.25ex}(E_{\alpha}+E_{\beta})\delta_{\alpha\alpha'}+\bra{\alpha\beta}\lambda V\ket{\alpha'\beta}+\frac{1}{2}\hspace{-0.5ex}\sum_{\alpha''\beta''}\hspace{-0.5ex}\bra{\alpha\beta} \lambda V\ket{\alpha''\beta''}\bra{\alpha''\beta''}\lambda V\ket{\alpha'\beta}\hspace{-0.5ex}\left(\frac{1}{\omega_{\alpha\alpha''}+\omega_{\beta\beta''}}+\frac{1}{\omega_{\beta\beta''}+\omega_{\alpha'\alpha''}}\right)\hspace{-0.5ex}.
\end{equation}
Since we have fast and slow components $\omega_{\beta\beta''}\gg\omega_{\alpha\alpha''}$, we may expand $(\omega_{\alpha\alpha''}+\omega_{\beta\beta''})^{-1}\approx \omega_{\beta\beta''}^{-1}(1-\omega_{\alpha\alpha''}/\omega_{\beta\beta''})$ in Eq.~\eqref{effective_ham}, which allows us to get an effective Hamiltonian that only acts on the slow degrees of freedom
\begin{equation}
\label{effective_ham_bis}
H^{\beta}_{\text{eff}}\approx P_{\beta}(H_0+\lambda V)P_{\beta}+\lambda^2\sum_{\beta''}P_{\beta}\frac{V\ket{\beta''}\bra{\beta''} V}{\omega_{\beta\beta''}}P_{\beta}.
\end{equation}
In this manuscript, we  make use of this canonical transformation to adiabatically eliminate the fast degrees of freedom of the electrons localized around a Nitrogen-vacancy impurity in diamond, and obtain an effective Hamiltonian for the slow $^{14}\text{N}$ nuclei.

{\it Effective nuclear  Hamiltonian for the static regime.-} According to the above discussion, one should first identify the fast and slow degrees of freedom of the Hamiltonian in Eqs.~\eqref{local_hamiltonian}-\eqref{dipole_hamiltonian}. Regarding the parameters in Table~\ref{o_magnitude}, together with the energy levels in Fig.~\ref{effective_coupling}{\bf(b)}, it is clear that the fast degrees of freedom correspond to electronic spin flips $\beta=\{m_1,m_2\}$, whereas the slow degrees of freedom involve the nuclei $\alpha=\{M_1,M_2\}$. Besides, the weak perturbation corresponds to the hyperfine interaction, which couples different electronic manifolds. We are interested in the dynamics within the electronic ground-state manifold  $\beta=\{0,0\}$, where the Schrieffer-Wolff transformation allow us to get the following Hamiltonian

\begin{equation}
\bra{\alpha}H_{\rm eff}^{00}\ket{\alpha'}=-\sum_{j}P_j(I_j^z)^2+\frac{1}{2}\sum_{\alpha''\beta''}\bra{00,\alpha}\sum_jH_j^{\text{(e-n)}}\ket{\beta''\alpha''}\bra{\beta''\alpha''}\sum_kH_k^{\text{(e-n)}}\ket{00,\alpha'}\left(\frac{1}{\omega_{\alpha\alpha''}+\omega_{\beta\beta''}}+\frac{1}{\omega_{\beta\beta''}+\omega_{\alpha'\alpha''}}\right),
\end{equation}
where the energy zero  is shifted to $E_0=\sum_j \frac{2}{3} (P_j-D_j)$, and one has to sum over all the electronic states $\{\ket{\beta''}\}$ on the first-excited manifold in Fig.~\ref{effective_coupling}{\bf(b)}. There are thus two types of channels of virtual electron spin flips, and  these correspond to the symmetric  $\ket{{\rm S}_{\pm}}=(\ket{0,\pm1}_{\rm e}+\ket{\pm1,0}_{\rm e})/\sqrt{2}$, and anti-symmetric  $\ket{{\rm A}_{\pm}}=(\ket{0,\pm1}_{\rm e}-\ket{\pm1,0}_{\rm e})/\sqrt{2}$ combinations. Due to the dipole-dipole interaction, these channels are split in energies by $E_{{\rm s}}-E_{{\rm a}}=2J_{12}$, which is the key ingredient that allow us to obtain a non-vanishing interaction between the nuclear spins. By expanding to leading order for $\beta''=S_{\pm},A_{\pm}$, we obtain $(\omega_{\alpha\alpha''}+\omega_{00\beta''})^{-1}\approx \omega_{00\beta''}^{-1}\approx D^{-1}(1\mp J_{12}/D)$ for the symmetric/anti-symmetric states. By resuming the expression above, one finally arrives to the effective nuclear Hamiltonian, which reads
\begin{equation}
H_{\rm eff}^{00}=-\sum_{j}P_j(I_j^z)^2+J_{\rm eff}^{\rm xx}(I_1^+I_2^-+I_1^-I_2^+),
\end{equation}
where $J_{\rm eff}^{\rm xx}=2A^{\bot}_1A^{\bot}_2J_{12}/D^2$, and we have a negligible local energy shift for $(A_{j}^{\bot})^2/D\ll D$. This is precisely the spin-spin Hamiltonian  $H_{\rm eff}^{\rm xx}$ described in Eq.~\eqref{eff_static}, which leads to the exchange of spin excitations between the nuclei. Let us finally remark that in this derivation we have assumed that there is no inhomogeneous broadening $D_1=D_2$, which may  modify the results. However, since the effective interactions are so weak $J_{\rm eff}^{\rm xx}\approx 0.1$Hz, there is no point of being more rigorous at this point. In the next section, we shall treat the possible effects of inhomogeneous broadening for the driven interactions in detail.

{\it Effective nuclear  Hamiltonian for the driven regime.-} In this part of the Appendix, we discuss the RWA leading to Eq.~\eqref{driven_hamiltonian}, and the Schrieffer-Wolff transformation to the effective nuclear Hamiltonian in Eq.~\eqref{ising}.

{\it a) Driven Hamiltonian.-} Let us rewrite the total driven Hamiltonian in Eqs.~\eqref{local_hamiltonian},\eqref{dipole_hamiltonian}, and~\eqref{driving} as follows $H=H_{0,1}+H_{0,2}$, where 
\begin{equation}
\label{rearrange}
\begin{split}
H_{0,1}&=\hspace{-0.5ex}\sum_{j}D_j(S_j^z)^2\hspace{-0.5ex}+g_{\text{e}}\mu_{\text{B}}B\cos\theta_jS_j^z-P_j(I_j^z)^2\hspace{-0.5ex}-g_{\text{n}}\mu_{\text{N}}B\cos\theta_jI_j^z,\\
H_{0,2}&=\sum_{j} \half B\sin\theta_j(g_{\text{e}}\mu_{\text{B}}\ee^{-\ii\varphi_j}S_j^+-g_{\rm n}\mu_{\text{N}}\ee^{-\ii\varphi_j}I_j^++\text{H.c.})+\sum_jH^{\text{(e-n)}}_j+H_{12}^{\text{(e-e)}}+H_{\rm d}(t),
\end{split}
\end{equation}
where we again shifted the energy zero  to $E_0=\sum_j \frac{2}{3} (P_j-D_j)$, and we have introduced the relative orientation of the NV centers $(\theta_j,\varphi_j)$ with respect to the applied magnetic fields.
In the interaction picture $H_{0,2}(t)=\ee^{\ii H_{0,1}t}H_{0,2}\ee^{-\ii H_{0,1}t}$, one can neglect rapidly oscillating terms by a rotating wave approximation (RWA), 
which leads us to 
\begin{equation}
\begin{split}
H^{\rm rwa}_{0,2}&\approx\sum_{j}\left(\half \Omega_{\text{e}}\sigma^x_j+\half\Omega_{\text{n}}\tau^x_j\right)+2J_{12}S_1^zS_2^z+\sum_{j}A_j^{\shortparallel}S_j^zI_j^z,
\end{split}
\end{equation}
where we define  $B_j=B\cos \theta_j$. This RWA is justified when $D_j\gg g_{\text{e}}\mu_{\text{B}}B_j\gg \Omega_{\rm e}, A_{j}^{\bot}, P_j\gg g_{\text{n}}\mu_{\text{B}}B\gg\Omega_{\rm n}$, which is clearly fulfilled for the parameters shown in Table~\ref{o_magnitude}. Additionally, the inhomogeneous splitting allows us to neglect the transverse part of the magnetic dipole coupling between the electron spins $J_{12}\ll g_{\text{e}}\mu_{\text{B}}|B_1-B_2|$. According to Table~\ref{o_magnitude},  one can neglect these terms $A^{\bot}_j/D_j\sim 10^{-3}$, $g_{\text{e}}\mu_{\text{B}}B/D_j\sim 10^{-2}$, $\Omega_{\rm e}/g_{\text{e}}\mu_{\text{B}}B\sim 10^{-1}$, $g_{\text{n}}\mu_{\text{N}}B/P_j\sim 10^{-3}$,  $\Omega_{\rm n}/g_{\text{n}}\mu_{\text{N}}B\sim 10^{-1}$. Besides, for two NV centers oriented along different axes, $\theta_1-\theta_2\sim\mathcal{O}(\pi)$, we can neglect the transverse dipole-dipole coupling  $J_{12}/ g_{\text{e}}\mu_{\text{B}}|B_1-B_2|\sim 10^{-3}$.

 In order to confirm the validity of these approximations, we must compare the dynamics of both Hamiltonians, $H_{0,2}(t),H_{0,2}^{\rm rwa}$. Let us note that the full time-dependent Hamiltonian $H_{0,2}(t)=\ee^{\ii H_{0,1}t}H_{0,2}\ee^{-\ii H_{0,1}t}$ contains very different time-scales,  ranging from  $ns$ to $ms$. To reproduce the dynamics faithfully, one sets the numerical integration time-step to the smallest time-scale, $ns$. For such a small time-step, prohibitively large integration times are required in order to reach the $ms$-regime where the nuclear spin-spin interaction effects become visible. Nonetheless, to test the accuracy of the RWA, it suffices to study   $t\in[0,2\pi/A_j^{\shortparallel}]$, which lies in the $\mu s$-range. In Fig.~\ref{fid}{\bf(a)}, we compare both predictions numerically  for the initial state  $\ket{\psi_0}=\ket{{\rm --}}_{\rm e}\otimes\ket{{\rm -+}}_{\rm n}$, namely
 \begin{equation}
 \begin{split}
 \langle \tau_j^x(t)\rangle_{\rm rwa}&=\bra{\psi_0}\ee^{\ii H_{0,2}^{\rm rwa}t}\tau_j^x\ee^{-\ii H_{0,2}^{\rm rwa}t}\ket{\psi_0},\hspace{2ex}\langle \tau_j^x(t)\rangle_{\rm exact}=\bra{\psi_0}\ee^{\ii\int_0^t {\rm d}t' H_{0,2}(t')}\tau_j^x\ee^{-\ii\int_0^t {\rm d}t'' H_{0,2}(t'')}	\ket{\psi_0}.
 \end{split}
 \end{equation}
 Since there is no refocusing echo pulse, the nuclear spin dynamics should be dominated by the Rabi oscillations caused by the term $-\fourth A_j^{\shortparallel}\tau_j^z$
 in Eq.~\eqref{ising}. As observed in Fig.~\ref{fid}{\bf(a)}, these neat $\mu s$ Rabi flops display a perfect agreement between the exact Hamiltonian and the RWA approximation.

\begin{figure}
	\centering
	\includegraphics[width=0.8\columnwidth]{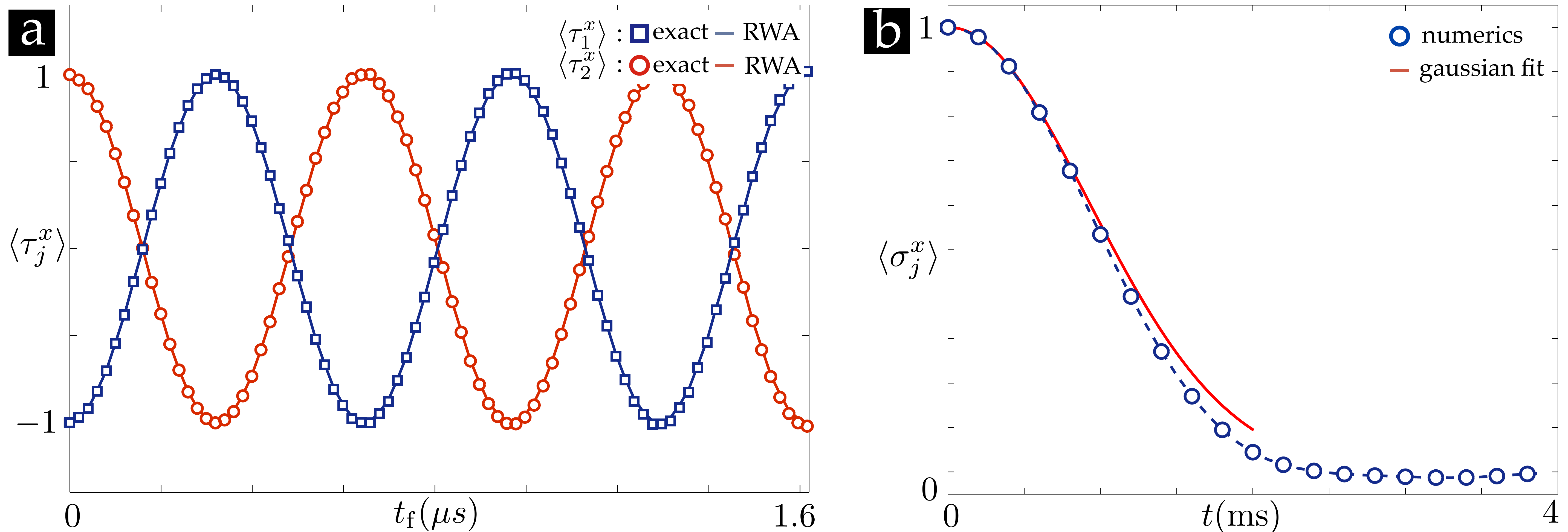}
	\caption{ {\bf Rotating wave approximation and free induction decay}. (a)  { Accuracy of the rotating wave approximation}: Dynamics of the expectation values  $\langle \tau_j^x(t)\rangle_{\rm rwa}$ (squares, circles), as solved by numerical exponentiation, and $\langle \tau_j^x(t)\rangle_{\rm exact}$ (solid lines), as solved numerically by a fourth order Runge-Kutta method.  (b){ Free induction decay due to magnetic noise:} Numerical simulation of the stochastic dynamics of the free induction decay  $\langle \tau_x(t)\rangle$ (circles), for the initial state $\ket{\Psi_0}=(\ket{0}_{\rm e}+\ket{-1}_{\rm e})/\sqrt{2},$ after averaging for $N_{\rm it}=5\cdot 10^3$ trajectories of the random process~\eqref{update}. The red line corresponds to a gaussian fit $\langle \tau_x(t)\rangle_{\rm fit}\propto {\rm exp}({-\half b_{\rm fit}^2 t^2})$, where $b_{\rm fit}=1.09$kHz. }
	\label{fid}
\end{figure}

{\it b) Effective nuclear Hamiltonian:} According to the preceding discussion, we  shall consider directly the RWA Hamiltonian in Eq.~\eqref{driven_hamiltonian}, where  the set of Pauli matrices is defined as follows
\begin{equation}
\begin{split}
\sigma_j^{z}&\hspace{-0.5ex}=\hspace{-0.5ex}\ket{0_j}_{\rm e}\bra{0_j}_{\rm e}\hspace{-0.25ex}-\hspace{-0.25ex}\ket{{\rm-1}_j}_{\rm e}\bra{{\rm -1}_j}_{\rm e}, \hspace{0.5ex}\sigma_j^{x}\hspace{-0.5ex}=\hspace{-0.5ex}\ket{{\rm-1}_j}_{\rm e}\bra{0_j}_{\rm e}\hspace{-0.25ex}+\hspace{-0.25ex}\text{H.c.} \\
\tau_j^{z}&\hspace{-0.5ex}=\hspace{-0.5ex}\ket{0_j}_{\rm n}\bra{0_j}_{\rm n}\hspace{-0.25ex}-\hspace{-0.25ex}\ket{{\rm-1}_j}_{\rm n}\bra{\!{\rm -1}_j}_{\rm n}, \hspace{0.5ex} \tau_j^{x}\hspace{-0.5ex}=\hspace{-0.25ex}\ket{{\rm-1}_j}_{\rm n}\bra{0_j}_{\rm n}\hspace{-0.25ex}+\hspace{-0.25ex}\text{H.c.}
\end{split}
\end{equation}
and the sub-indexes $\ket{\hspace{1ex}}_{\rm e},\ket{\hspace{1ex}}_{\rm n}$ indicate the electronic or nuclear origin of the spin state. 
In this two-level approximation, the spin-1 operators become $S_j^{z}=\ket{1_j}_{\rm e}\bra{1_j}_{\rm e}+\half(\sigma_j^z-\mathbb{I}_2)$, and $I_j^{z}=\ket{1_j}_{\rm n}\bra{1_j}_{\rm n}+\half(\tau_j^z-\mathbb{I}_2)$. Accordingly, the electronic and nuclear levels $m_j=1,M_j=1$ decouple, and one may write the following driven pseudospin-1/2 Hamiltonian 
\begin{equation}
\label{2_level_driven_hamiltonian}
\begin{split}
H_0&=\hspace{-0.5ex}\sum_{j}\left(\half \Omega_{\text{e}}\sigma^x_j+\half\Omega_{\text{n}}\tau^x_j\right)+\half J_{12}(\sigma_1^z-\mathbb{I}_2)(\sigma_2^z-\mathbb{I}_2),\hspace{2ex} H_1=\sum_{j}\textstyle{\frac{1}{4}}A_j^{\shortparallel}(\sigma_j^z-\mathbb{I}_2)(\tau_j^z-\mathbb{I}_2),
\end{split}
\end{equation}
Once  the pseudospin-1/2 Hamiltonian in Eq.~\eqref{2_level_driven_hamiltonian} has been derived,
we adiabatically eliminate the fast electronic degrees of freedom from the slow nuclear dynamics by a Schrieffer-Wolff  transformation. We shall make use of Eq.~\eqref{effective_ham_bis}, where we identify $\beta=\{-,-\}$ as the lowest-energy manifold (see Fig.~\ref{effective_coupling}{\bf(d)}), $\beta''={\rm S,A}$ as the symmetric/anti-symmetric excited manifolds 
\begin{equation}
\begin{split}
\ket{\rm S}=\frac{1}{\sqrt{1+(1+\xi)^2}}\big((1+\xi)\ket{+-}_{\rm e}+\ket{-+}_{\rm e}\big), \hspace{2ex}\ket{\rm A}=\frac{1}{\sqrt{1+(1-\xi)^2}}\big((1-\xi)\ket{+-}_{\rm e}-\ket{-+}_{\rm e}\big),
\end{split}
\end{equation}
where we have introduced $\ket{\pm}_{\rm e}=(\ket{0}_{\rm e}\pm\ket{\rm{-1}}_{\rm e})/\sqrt{2}$, and a parameter $\xi\ll1$ quantifying the inhomogeneous broadening  $A_1^{\shortparallel}\neq A_2^{\shortparallel}$. These levels  are split in energies by $2J(\xi)$, where
\begin{equation} 
J(\xi)=\textstyle{\frac{J_{12}}{2}}\sqrt{1+\xi^2},\hspace{2ex} \xi=\frac{2}{\Omega_eJ_{12}}\left[(A_2^{\shortparallel})^2-(A_1^{\shortparallel})^2\right],
\end{equation} 
 To second order in the hyperfine coupling, we are able to derive
\begin{equation}
H_{\text{eff}}^{\text{-}\text{-}}\hspace{-0.5ex}=\hspace{-0.5ex}\sum_{j}\hspace{-0.5ex}(\Omega_{\text{n}}\tau_j^x+\frac{1}{4} A_j^{\shortparallel}\tau_j^z)-\frac{1}{16\Omega_{\text{eS}}}\bra{{\rm --}}\sum_{j} A_j^{\shortparallel}\sigma_j^z\tau_j^z\ket{{\rm S}}\hspace{-0.5ex}\bra{{\rm S}}\sum_{k}  A_k^{\shortparallel}\sigma_k^z\tau_k^z\ket{{\rm --}} -\frac{1}{16\Omega_{\text{eA}}}\bra{{\rm --}}\sum_{j}  A_j^{\shortparallel}\sigma_j^z\tau_j^z\ket{{\rm A}}\hspace{-0.5ex}\bra{{\rm A}}\sum_{k} A_k^{\shortparallel}\sigma_k^z\tau_k^z\ket{{\rm --}}\hspace{-0.25ex},
\end{equation}
where we have introduced the following energy differences
\begin{equation}
\begin{split}
\textstyle{
\Omega_{\text{eS}}=\Omega_{\text{e}}\left[1+\left(\frac{A_1^{\shortparallel}}{\Omega_e}\right)^2+\left(\frac{A_2^{\shortparallel}}{\Omega_e}\right)^2\right]+J(\xi),} \hspace{2ex}\textstyle{
\Omega_{\text{eA}}=\Omega_{\text{e}}\left[1+\left(\frac{A_1^{\shortparallel}}{\Omega_e}\right)^2+\left(\frac{A_2^{\shortparallel}}{\Omega_e}\right)^2\right]-J(\xi).}
\end{split}
\end{equation}
By computing the corresponding matrix elements, together with a Taylor expansion for $\Omega_{\text{e}}\gg A_{j}^{\shortparallel},J_{12}$, and $1\gg\xi$, we find the following expression for the  followig nuclear spin Hamiltonian, which is precisely Eq.~\eqref{ising} in the main text,
\begin{equation}
H_{\text{eff}}^{\text{-}\text{-}}\hspace{-0.5ex}=J^{\rm zz}_{\text{eff}}\tau_1^z\tau_2^z+\hspace{-0.5ex}\sum_j\hspace{-0.5ex}\Omega_{\text{n}}\tau_j^x-\frac{1}{4} A_j^{\shortparallel}\tau_j^z,\hspace{0.5ex} J^{\rm zz}_{\text{eff}}=\hspace{-0.5ex}\frac{-A_1^{\shortparallel}A_2^{\shortparallel}}{8\Omega_e}\hspace{-0.5ex}\left(\hspace{-0.5ex}\frac{J_{12}}{\Omega_{\text{e}}}+2\xi\hspace{-0.5ex}\right)\hspace{-0.5ex},
\end{equation}

\subsection{ Decoherence and effective decoupling by continuous microwave driving}

In order to perform quantum-information tasks in a solid-state device, the effects of the system-environment coupling must be carefully addressed. In contrast to cold-atom platforms, the environment in a solid is rather complex since the spins may couple to a wide variety of excitations. In the case of NV centers, whose energy levels lie deep in the band gap of diamond, the major source of noise is the coupling to the spins of different impurities, rather than to electronic or vibronic excitations. Accordingly,   
one should consider the effects of a spin bath on the coherent features of the electron/nuclear spin of the NV center. 

{\it Phenomenological magnetic noise model.-} The problem of a central spin coupled to an ensemble of bath spins has been studied since the early days of nuclear magnetic resonance~\cite{anderson_weiss}, and depending on the particular nature of the spin bath can be an intricate many-body problem.  For type Ib diamond, the bath consists of the electronic spins of $^{14}$N impurities, the so-called P1 centers, randomly distributed through the sample. The dipolar coupling of the P1 centers to the NV electron spin gives rise to a pure dephasing which can be treated by mean-field theories~\cite{adjustable_spin_bath_nv}. Conversely, for ultrapure type IIa diamond, it is the nuclear spin of $^{13}$C isotopes which yields the dephasing of the NV center via the hyperfine electron-nuclei coupling~\cite{c13_environment}. Interestingly, the correlations of this nuclear-spin environment must  be accounted in order to reproduce the short-time dynamics of the system. In this work, we follow a phenomenological approach rather than a microscopic one, where the magnetic noise is modeled by a random fluctuation of the resonance frequencies associated to the electron/nuclear spins. This model captures the whole dynamics of type Ib diamond~\cite{dynamical_decoupling_nv}, and is expected to describe faithfully the long-time dynamics of type IIa diamond ($t_{\rm f}\sim$ms), where the non-Markovian aspects of the environment should not have an important effect.

 The flip-flop interactions between the NV and the bath spins can be safely neglected due to their utterly different energy scales. Therefore,  we consider that the collective effect of the spin bath is to shift the resonance energies of the NV center (i.e. pure dephasing), which can be modeled by an effective local magnetic field. In order to account for the spin-bath dynamics, this magnetic field is treated as a stochastic process~\cite{dynamical_decoupling_nv}.
Hence,  the NV centers are described by a stochastic Hamiltonian 
\begin{equation}
\label{stochastic_ham}
H(\{b_j(t),B_j(t)\})=\sum_j \left(H_j^{\rm (e)}+H_j^{\rm (n)}+H_j^{\rm (e-n)}\right)+H_{12}^{\rm e-e}+H_{\rm d}(t)+ H_{\rm noise}, \hspace{2ex}H_{\rm noise}=\sum_j \left( b_{j}(t)S_j^z+ B_{j}(t)I_j^z\right),
\end{equation}
where $b_{j}(t), B_{j}(t)$ are the random processes describing the fluctuating resonance frequencies. Whereas the electron and nuclear spins of the NV center follow the unitary dynamics dictated by the Schr\"{o}dinger equation $\ii\partial_t\ket{\Psi}=H(\{b_j(t),B_j(t)\})\ket{\Psi}$, the fluctuating fields evolve according to the so-called Langevin equation $d_t X(t)=A(X(t),t)X(t)+\sqrt{D(X(t),t)}\Gamma(t)$, where $X(t)=\{b_j(t),B_j(t)\}$ is the random process, $A(x,t)$, and $D(x,t)>0$, are smooth functions, and $\Gamma(t)$ is a Gaussian zero-mean noise (i.e. $\langle \Gamma(t)\rangle=0, \langle \Gamma(t)\Gamma(t')\rangle=\delta(t-t')$)~\cite{random_processes}.  Due to the large number of spins conforming the bath, these random processes can be argued to be {\it Gaussian} by means of the central limit theorem. Besides, when the back-action of the system is small~\cite{dynamical_decoupling_nv}, or we are interested in the long-time dynamics, the process can be treated as {\it Markovian} and {\it stationary}.  In this case, one obtains a Langevin equation with  $A(x,t)=-x/\tau$, and $D(x,t)=c$, where $\tau$ is the relaxation time of the process, and $c$ the diffusion speed. The solution to this equation  can be obtained explicitly, and is known as the {\it Ornstein-Uhlenbeck} (OU) process~\cite{ou_process}.  In the  case of zero-mean random magnetic fields, the auto-covariances are
\begin{equation}
\label{stochastic_ham}
\begin{split}
\langle b_j(t)b_j(0)\rangle = b_j^2\ee^{-r_j t}, \hspace{1ex} \langle B_j(t)B_j(0)\rangle =B_j^2\ee^{-R_j t},\\
\end{split}
\end{equation}
where $b_j^2=\half c_{b_j}\tau_{b_j},B_j^2=\half c_{B_j}\tau_{B_j}$ represent the variances of the zero-mean gaussian distributions, and $r_j=1/\tau_{b_j},R_j=1/\tau_{B_j}$ the inverse of their relaxation times. These auto-correlations lead to a Lorentzian spectral density, which contains a white-noise region at low frequencies, and $1/f^2$-noise region at larger frequencies. Interestingly enough, the dynamics of the OU process can be given explicitly~\cite{ou_process} , and numerical integration of the Langeving equation is not required. In fact, for any discretization ${\rm d}t>0$, one finds the following exact update formula
\begin{equation}
\label{update}
X(t+{\rm d}t)=X(t)\ee^{-t/\tau}+\sqrt{\frac{c\tau}{2}(1-\ee^{-2{\rm d}t/\tau})}n,
\end{equation}
where $n$ is a zero-mean unit-variance gaussian random variable which is time uncorrelated. In order to solve the whole stochastic quantum dynamics in Eq.~\eqref{stochastic_ham}, we discretize the time interval in $M$ time-steps, $t_m=m \hspace{0.1ex} {\rm d} t\in[0,t_{\rm f}]$, where ${\rm d}t=t_{\rm f}/M$, and obtain the different values of the fluctuating magnetic fields $b_j(t_m),B_j(t_m)$ by employing the above formula~\eqref{update}. Then, we integrate numerically the stochastic Hamiltonian for the particular sampling of the random process  $s=\{b_j(t_m),B_j(t_m)\}$, and recover the expectation values $\langle\tau_j^x\rangle_{s}$. By repeating this procedure for $N_{\rm it}\gg 1$, one can perform the statistical average over the stochastic noise, $\langle\tau_j^x\rangle=\frac{1}{N_{\rm it}}\sum_{s}\langle\tau_j^x\rangle_{s}$, and thus study the effects of the decoherence. 

To illustrate the physics of this phenomenological model, let us consider the simpler situation of a single NV electron spin. We consider the decoherence of a Ramsey experiment, where the initial state corresponds to $\ket{\Psi_0}=(\ket{0}_{\rm e}+\ket{-1}_{\rm e})/\sqrt{2},$ and we measure the so-called free induction decay (FID) due to the noise after a certain time $t$, $\langle \sigma^x(t)\rangle$. In Fig.~\ref{fid}{\bf(b)}, we represent the time evolution of the FID derived from the numerical solution of the stochastic Hamiltonian $H(b(t))=b(t)S^z$, where $b(t)$ is a OU process with $b=1$kHz, and we have averaged over $N_{\rm it}=5\cdot 10^3$ samplings of the random process. Due to the magnetic noise, the free induction decay follows a gaussian decay law $\langle \tau_x(t)\rangle\propto {\rm exp}({-\half b^2 t^2})$, which allows us to identify the dephasing time as $T_{\rm 2,e}=1/b\approx 1$ms. Therefore, we observe how the phenomenological noise model allows us to study the decoherence effects for different dephasing rates, which has been used in the main text (Fig.~\ref{nuclear_gate}{\bf(e)}).

{\it Effective decoupling mechanisms.-} An advantage of the phenomenological noise models is that they allow a neat understanding of the effects of decoherence, together with possible strategies to overcome them. In the particular case of the driven Hamiltonian~\eqref{driven_hamiltonian}, it is easy to observe that the fluctuation of the electronic resonance frequencies, $H=\sum_j b_j(t)S_j^z$, tries to induce transitions between the  energy manifolds of Fig.~\ref{effective_coupling}{\bf(d)}, namely $\ket{+_j}_{\rm e}\leftrightarrow\ket{-_j}_{\rm e}$. However, since these states now have a huge energy difference given by the Rabi frequency of the driving $\Omega_{\rm e}$, these transitions are non-resonant and thus partially suppressed. In fact, the electron magnetic noise can only couple to the nuclei via second order processes. The leading order contribution comes from the coupling to the hyperfine channel, and gives rise to $H_{\rm eff}\approx\sum_j (b_jA_j^{\shortparallel}/\Omega_{\rm e})\tau_j^z$, which is partially suppressed for the regime considered in this work $b_j,A_j\ll\Omega_{\rm e}$. Now, one has to compare this new term to the nuclear driving, and since $(b_jA_j^{\shortparallel}/\Omega_{\rm e})\ll \Omega_{\rm n}$, we get an additional decoupling mechanism. Qualitatively, one can argue that the effects of the noise give rise to a  small second-order fluctuation of the nuclear driving $H_{\rm eff}\approx\sum_j\Omega_{\rm n}\big(1+\half(b_jA_j^{\shortparallel}/\Omega_{\rm e})^2/\Omega_{\rm n}^2\big)\tau_j^x$. In this expression, one observes the two-fold role of the microwave driving $\Omega_{\rm e}$. On the one hand, $\Omega_{\rm e}$ must be small enough so as to increase the effective nuclear interaction. On the other hand,  $\Omega_{\rm e}$ must be big enough so as to provide an effective decoupling from the electronic noise. Therefore, one must find a compromise between the two, such as that presented for the parameters in Table~\ref{o_magnitude}. With respect to the additional decoupling due to the driving of the nuclei, $\Omega_{\rm n}$ must be as big as possible. By increasing the external magnetic fields beyond $B\approx 500$G, where the levels $m_j=0,-1$ become degenerate, one could raise the nuclear driving strength, and thus increment the efficiency of the decoupling.

\end{widetext}

\end{document}